\documentclass[aps,onecolumn, notitlepage, floatfix, nofootinbib, amsmath, letterpaper,amssymb, showpacs,showkeys,superscriptaddress,nobalancelastpage,numerical]{revtex4-1}

\usepackage{amsmath, amssymb, graphics, setspace}

\usepackage{graphicx}
\usepackage{amssymb}
\usepackage{tabu,amsmath}
\usepackage{setspace}
\usepackage{epstopdf}
\usepackage{slashed}

\newcommand{\mathsym}[1]{{}}
\newcommand{\unicode}[1]{{}}

\usepackage{verbatim}

\usepackage{mathpazo}

\usepackage{amsbsy}

\usepackage{url}

\usepackage[makeroom]{cancel}

\usepackage{mathpazo}

\usepackage{geometry}

\usepackage{slashed}
\usepackage{xcolor}

\usepackage{multirow}
\usepackage{rotating}
\usepackage[normalem]{ulem}

\usepackage{wrapfig}

\usepackage[font=small,labelfont=bf]{caption}

\newcommand{\ignore}[1]{}
\newcommand{\be}{\begin{equation}} \newcommand{\ee}{\end{equation}}

\def\ba#1\ea{\begin{align}#1\end{align}}
\newcommand{\bit}{\begin{itemize}}
\newcommand{\eit}{\end{itemize}}
\newcommand{\im}{\item}

\newcommand{\nnn}{\nn \\ &}

\newcommand{\nn}{\nonumber} 
\newcommand{\ra}{\rightarrow}

\renewcommand{\a}{\alpha} \renewcommand{\b}{\beta}

\def\slasha#1{\setbox0=\hbox{$#1$}#1\hskip-\wd0\hbox 
to\wd0{\hss\sl/\/\hss}}
\def\slashb#1{\setbox0=\hbox{$#1$}#1\hskip-\wd0\dimen0=5pt\advance
        \dimen0 by-\ht0\advance\dimen0 by\dp0\lower0.5\dimen0\hbox
          to\wd0{\hss\sl/\/\hss}}

\input{epsf}

\begin{document}

\title{Quantum tomography for collider physics: Illustrations with lepton pair production}
\author{John C. Martens}
\email{martens@ku.edu, daniel.tapia.takaki@cern.ch}
\affiliation{Department of Physics and Astronomy, The University of Kansas, Lawrence, KS 66045, USA\vspace{0ex}}
\author{John P. Ralston}
\affiliation{Department of Physics and Astronomy, The University of Kansas, Lawrence, KS 66045, USA\vspace{0ex}}
\author{J. D. Tapia Takaki}
\affiliation{Department of Physics and Astronomy, The University of Kansas, Lawrence, KS 66045, USA\vspace{0ex}}
\begin{abstract}
{\it Abstract:} Quantum tomography is a method to experimentally extract all that is observable about
a quantum mechanical system. We introduce quantum tomography to collider physics with the illustration of the angular distribution of lepton pairs. The tomographic method bypasses much of the field-theoretic formalism to concentrate on what can be observed with experimental data, and how to characterize the data. We provide a practical, experimentally-driven guide to model-independent analysis using density matrices at every step. Comparison with traditional methods of analyzing angular correlations of inclusive reactions finds many advantages in the tomographic method, which include manifest Lorentz covariance, direct incorporation of positivity constraints, exhaustively complete polarization information, and new invariants free from frame conventions. For example, experimental data can determine the {\it entanglement entropy} of the production process, which is a model-independent invariant that measures the degree of coherence of the subprocess. We give reproducible numerical examples and provide a supplemental standalone computer code that implements the procedure. We also highlight a property of {\it complex positivity} that guarantees in a least-squares type fit that a local minimum of a $\chi^{2}$ statistic will be a global minimum: There are no isolated local minima. This property with an automated implementation of positivity promises to mitigate issues relating to multiple minima and convention-dependence that have been problematic in previous work on angular distributions.
\end{abstract}
\keywords{polarization, entanglement, entropy, inclusive production, dilepton, quarkonium, angular distribution, data analysis method, intermediate state, density matrix\vspace{-5ex}}
\maketitle

\section{Introduction}

{\it Tomography} builds up higher dimensional objects from lower dimensional projections. {\it Quantum tomography} \cite{Fano} is a strategy to reconstruct all that can be observed about a quantum physical system. After becoming a focal point of quantum computing, quantum tomography has recently been applied in a variety of domains \cite{2004PhRvA..69d2108H, quantcomp00, quantcomp10, 2016JETPL.104..510S, 2010PhRvL.105o0401G, 2017arXiv170208751B, 2015NJPh...17d3063F, 2012arXiv1204.5936A}. 

The method of quantum tomography uses a known ``probe'' to explore an unknown system. Data is related directly to matrix elements, with minimal model dependence and optimal efficiency. 

Collider physics is conventionally set up in a framework of unobservable and model-dependent scattering amplitudes. In quantum tomography these unobservable features are skipped to deal directly with observables. The unknown system is parameterized by a certain density matrix $\rho(X)$, which is model-independent. The probe is described by a known density matrix $\rho(probe).$ The matrices are represented by {\it numbers} generated and fit to experimental data, not {\it abstract operators.} Quantum mechanics predicts an experiment will measure $tr(\rho(probe) \cdot \rho(X))$, where $tr$ is the trace. In many cases $\rho(probe)$ is extremely simple: A $3\times 3$ matrix, say. {\it What will be observed is strictly limited by the dimension and symmetries of the probe.} The powerful efficiency of quantum tomography comes from exploiting the probe's simplicity in the first steps. The description never involves more variables than will actually be measured. 

We illustrate the advantages of quantum tomography with inclusive lepton-pair production. It is a relatively mature subject chosen for its pedagogical convenience.  Despite the maturity of the subject, we discover new things. For example, the puzzling plethora of plethora of ad hoc invariant quantities is completely cleared up. We also find new ways to assist experimental data analysis. Positivity is a central issue overlooked in the literature, which we show how to control. Moreover, the tomography procedure carries over straightforwardly to many final states, including the inclusive production of charmonium, bottomonium, dijets, including boosted tops, $HH$, $W^+ W^-$, $ZZ$ \cite{Abelev:2011md, Aad:2016izn, Melnitchouk:2011tq, Brambilla:2010cs, Aaltonen:2011nr, Chatrchyan:2012ty, Chatrchyan:2013yna,Chatrchyan:2013cla, Chatrchyan:2012woa, Aaij:2013nlm, Aaij:2014qea, Mcclellan:2016cul, Peng:2014hta, Peng:2015spa, Stirling:2012zt, Han:2011vw, CMS:2012xwa, Khachatryan:2015qpa, Cheung:2017loo, Cheung:2017osx, Kang:2014pya, Cervera-Lierta:2017tdt, Kharzeev:2017qzs}.  
Our practical guide to analyzing experimental data uses density matrices at each step and circumvents the more elaborate traditional theoretical formalism. We concentrate on making tools available to experimentalists. We give a step-by-step guide where density matrices stand as definite arrays of {\it numbers}, bypassing unnecessary formalism. 

\section{The quantum tomography procedure applied to inclusive lepton pair production}

The tomography procedure reconstructs all that can be observed about a quantum physical system. For inclusive lepton pair production, what can be observed is the invariant mass distribution, the lepton pair angular distribution $dN/d\Omega$ and the polarization of the unknown intermediate state of the system, contained in $\rho(X)$.\footnote{ Polarization and spin are different concepts. The polarization (and density matrix of the unknown state) predicts the spin, while the spin cannot predict the density matrix.} In this section we reconstruct $dN/d\Omega$ and $\rho(X)$ from first principles using tomography. Structure functions and model-dependent assumptions about the intermediate state, common to the traditional formalism \cite{Terazawa:1974ci, Vasavada:1977ef, Mirkes:1992hu, Mirkes:1994eb, Mirkes:1994dp}, do not appear.

Expert readers, who are accustomed to seeing some of these formulas derived, might note that the method of derivation is particularly simple. The particular steps we do {\it not} follow are to be noted. That also explains why some of the relations we find seem to have been overlooked in the past.
 
\subsection{Kinematics}

Consider inclusive production of a lepton pair with 4-momenta $k$, $k'$ from the collision of two hadrons with 4-momenta $P_{A}$, $P_{B}$:  $$P_{A}P_{B} \ra \ell^{+}(k)\ell^{-}(k')+{\cal X},$$ where ${\cal X}$ and the final state lepton spins are unobserved and thus summed over. In the high energy limit  $k^{2} = k^{'2} =0$. 

Let the total pair momentum $Q=k+k'$. The azimuthal distribution of total pair momenta in the lab frame is isotropic. Lepton pair angular distributions are described in the pair rest-frame defined event-by-event.  In this frame the pair momenta are back-to-back and equal in magnitude. The frame orientation depends on the beam momenta and the pair total momentum. 

Defining momentum\footnote{This is a great advantage compared to making calculations with a complicated (and error prone) sequence of rotations and boosts.} observables via a Lorentz-covariant frame convention allows calculations to be done in {\it any} frame. In its rest frame the total pair momentum $Q^{\mu} = (\sqrt{Q^{2}}, \, \vec Q=0)$. A set of $xyz$ spatial axes in this frame will be defined by three 4-vectors $X^{\mu}, \, Y^{\mu}, \, Z^{\mu}$, satisfying \ba Q\cdot X=Q\cdot Y =Q\cdot Z=0. \label{Qdot} \ea The frame vectors being orthogonal implies \ba X\cdot Y=Y \cdot Z =X\cdot Z=0 \label{last}  \ea Taking $P_{A}$=(1, 0, 0, 1), $P_{B}$=(1, 0, 0, -1) (light-cone $\pm$ vectors),  a frame satisfying the relations of  Eq. \ref{Qdot} and Eq. \ref{last} is given by\footnote{We use $\epsilon^{0123}=1$. The mirror symmetry of $pp$ collisions also strongly supports a convention where the direction of the $Z$ axis is determined by the sign of the pair rapidity. The formulas shown do not include this detail.}  \ba \tilde Z^{\mu} &=  P_{A}^{\mu} Q\cdot P_{B}  -  P_{B} ^{\mu} Q\cdot  P_{A}; \nn \\ \tilde X^{\mu} &= Q^{\mu}- P_{A}^{\mu} {Q^{2}\over 2 Q\cdot P_{A}} - P_{B} ^{\mu} {Q^{2}\over 2 Q\cdot P_{B} }; \nn \\ \tilde  Y^{\mu}&= \epsilon^{\mu \nu \a \b}P_{ A\nu}P_{B \a}Q_{\b}. \nn \ea  These frame vectors define the Collins-Soper ($CS$) frame.\footnote{These expressions  simplify a more complicated convention that included finite mass effects in the original definition} 
The normalized frame vectors are \ba (X^{\mu}, \, Y^{\mu}, \, Z^{\mu}) =  ({\tilde X^{\mu} \over \sqrt{-\tilde X\cdot \tilde X} }, \, {\tilde Y^{\mu} \over \sqrt{-\tilde Y\cdot \tilde Y} }, \, {\tilde Z^{\mu} \over \sqrt{-\tilde Z\cdot \tilde Z} }) . \nn \ea

To analyze data for each event labeled $J$: \ba  \text{Compute} & \quad Q_{J} =k_{J}+k_{J}'; \quad \ell_{J} =k_{J}-k_{J}'; \quad (X_{J}^{\mu}, \, Y_{J}^{\mu}, \, Z_{J}^{\mu}) ; \nn \\ & \vec \ell_{XYZ, J} = ( X_{J}\cdot \ell_{J}, \, Y_{J}\cdot \ell_{J} , \, Z_{J}\cdot \ell_{J}); \nn \\  &\hat \ell_{J} =\ell_{XYZ, \, J}  /\sqrt{-\ell_{XYZ, \, J}\cdot \ell_{XYZ, \, J}}. \label{hatell} \ea

In fact, $\hat \ell_J = (\sin \theta \cos \phi, \,\sin \theta \sin \phi, \,\cos \theta  )_J,$ where $\theta$, $\phi$ are the polar and azimuthal angles of one (e.g. plus-charge) lepton in the rest frame of $Q$. The meaning of a "Lorentz invariant ${\it cos}$ $\theta$" is a scalar $Z_{\mu}(k-k')^{\mu}$ which becomes $\hat z \cdot \hat{(k - k')}$ in the rest frame of $Q.$

\subsection{The angular distribution, in terms of the probe and target density matrices}

The standard amplitude for inclusive production of a fermion- anti-fermion pair of spin $s$, $s'$ has a string of gamma-matrices contracted with final state spinors $v_{a}(k's')$, $\bar u_{b}(k, s)$. When the amplitude is squared, these factors appear bi-linearly, as in \ba  u_{a}(ks)\bar u_{a'}(ks) =(1/2)[( \slasha k+m)(1+\gamma_{5}\slasha s)]_{aa'}. \nn \ea Summing over unobserved $s$ and dropping $m \delta_{aa'}$, a form of density matrix appears: \ba \sum_{s} \,  u_{a}(ks)\bar u_{a'}(ks) \ra k_{\mu}\gamma_{aa'}^{\mu}. \nn \ea The Feynman rules for the density matrix of two relativistic final state fermions (or anti-fermions, or any combination) is a factor given by \ba \rho_{aa', \, bb'}(k, \, k') \ra \slasha k_{aa'} \slasha k'_{bb'}\label{clutter} \ea  This fundamental equality is not present in pure state quantum systems.
{\it There is no spinor corresponding to a fermion averaged over initial spins}, nor to a fermion summed over final spins.

As shown in the Appendix, the rest of the cross section appears in the target density matrix $\rho(X)$, which must have four indices to contract with the probe indices: \ba d\sigma \sim \sum_{aa'bb'} \, \rho_{aa', \, bb'}(k, \, k')\rho_{aa', \, bb'}(X) dLIPS = tr \left(\rho(k, \, k')\rho(X) \right) dLIPS, \label{LIPS} \ea where $dLIPS$ is the Lorentz invariant phase space.
 
Note that $ u_{a}(ks)\bar u_{a'}(ks)$ is not positive definite since the Dirac adjoint $ \bar u_{a'}(ks)= (u^{\dagger}(ks) \gamma_{0})_{a}$ has a factor of $ \gamma_{0}$, introduced by convention. Removing it, $\sum_{s} \, u_{a}(ks)u_{a'}^{\dagger}(ks)$ becomes positive by inspection. (Any matrix of the form $M \cdot M^{\dagger}$ has positive eigenvalues.) $\rho(k, \, k')$ as written is not normalized, because the Feynman rules shuffle spinor normalizations into overall factors. To make the arrow in Eq. \ref {clutter} into an equality, multiply on the right by $\gamma_{0}$ twice, and standardize the normalizations. The same steps applied to $\rho_{aa', \, bb'}(X)$ cancels the $\gamma_{0}$ factors. The result is that the probability to find two fermions has the fundamental quantum mechanical form\footnote{We remind the reader that the phase space factors $dLIPS$ originate in further organizational steps computing the quantum mechanical transition probability per volume per time, which afterwards restore the phase space factors. } $P(k, \,k')=tr(\rho(k, \, k')\rho(X))$. 

The left side of Eq. \ref{LIPS} is $d\sigma(k, \, k')$, the same as the joint probability $P(Q, \, \ell \big | \, init)$ where $init$ are the initial state variables. The phase space for two leptons converts as \ba k_{0}k_{0}' {d\sigma \over d^{3}k  d^{3}k'} = {d \sigma \over  d^{4}Q d\Omega } . \nn \ea We can write \ba P(Q, \, \ell \big | \, init)=P(\ell \big |Q, \, init)P(Q \big |init). \nn \ea Here $P(Q \big |init) = d\sigma/ d^{4}Q$, and $P(\ell \big |Q, \, init)=dN/d\Omega$ is the conditional probability to find $\ell$ given $Q$ and the initial state. This factorization is general and unrelated to one-boson exchange, parton model, or other considerations. Since $P(\ell \big |Q, \, init)$ is a probability, quantum mechanics predicts it is a trace: \ba & {dN\over d \Omega} =  {1\over \sigma}{d\sigma \over d\Omega} = P(\ell \big |Q, \, init)={3\over 4 \pi}tr(\rho(\ell) \rho(X)), \label{dsigma} \ea where $tr$ indicates the trace, $d \Omega = d cos \theta \cdot d \phi,$ and $\rho(\ell),$ the probe, is a $3\times 3$ matrix to be defined momentarily which depends only on the directions $\hat \ell_J$.  The target hadronic system is represented by $\rho(X)$. Since the probe $\rho(\ell)$ is a $3\times 3$ matrix, then $\rho(X)$ is a $3\times 3$ matrix of numbers. \footnote{This is a {\it more general} statement than enumerating ``structure functions''.}

The description has just been {\it reduced} from $\rho_{aa', \, bb'}(k, \, k')$, a Dirac tensor with $4^{4}$ possible matrix elements, to a $3\times 3$ Hermitian matrix with 8 independent elements, since $tr(\rho(\ell))=1$ is one condition. Equation \ref{dsigma} is the most general angular distribution that can be observed. It is valid for {\it like-sign and unlike sign pairs, and assumes no model for how the pairs are produced}. The Dirac form (and Dirac traces) is over-complicated, because describing every possible {\it exclusive} reaction for every possible in and out state is over-achieved in the formalism. 

\subsection{The probe matrix}

The probe matrix $\rho(\ell)$ is given by \ba \rho_{ij}(\ell) = {1+a\over 3}\delta_{ij} -a \hat \ell_{i}\hat \ell_{j} -\imath b \epsilon_{ijk}\hat \ell_{k}, \label{rholep} \ea which is derived in the Appendix. The Standard Model predicts only two parameters, $a$ and $b$. If on-shell lepton helicity is conserved (as in lowest order production by a minimally-coupled vector boson) then $a =1/2$ and $b = c_{A}c_{V}.$ The latter is not a prediction but a definition. If the production is parity-symmetric then $c_{A}=0$. {\it The only non-trivial prediction of the Standard Model is the value of $c_{A}c_{V}$}.  Lowest-order production by $Z$ bosons predicts $b=\sin^{2}\theta_{W}\sim 0.22$.

More generally, the probe matrix {\it itself} represents a reduced system that is unknown {\it a priori}. It should be determined experimentally. Consider the angular distribution of $e^{+}e^{-} \ra \mu^{+}\mu^{-}$. Let $\rho(e; \, \hat z)$ describe electrons with parameters $a_{e}, \, b_{e}$ colliding along the $z$ axis. Let $\rho(\mu; \, \hat \ell)$ describe muons with parameters $a_{\mu}, \, b_{\mu}$ emerging along direction $\hat \ell$. A short calculation using Eq. \ref{rholep} twice gives\footnote{This may be a new result, which goes beyond what is known from one-boson exchange with or without radiative corrections. The production details can only renormalize the parameters.} \ba  {3\over 4 \pi}tr \left(\rho(e; \, \hat z) \rho(\mu; \, \hat \ell) \right ) = &  {3\over 4 \pi} \left({1\over 3} + 2 b_{e}b_{\mu} \hat \ell \cdot \hat z+ a_{e}a_{\mu}((\hat z\cdot \hat \ell)^{2} -1/3) \right), \nn \\ &=   {3\over 4 \pi}\left({1\over 3} + 2 b_{e}b_{\mu}\cos\theta+  a_{e}a_{\mu} (\cos^{2}\theta-1/3) \right). \label{dist1}  \ea Fitting experimental data will give $ a_{e}a_{\mu}$ and  $b_{e}b_{\mu}$. If lepton universality is assumed the probe $\rho(\mu; \hat \ell)$ has measured the probe $\rho(e; \hat z)$. 

\subsection{How tomography works: $dN/d\Omega$ as a function of $\rho(\ell)$, $\rho(X)$} 

\label{sec:mirrror}

Let $\hat G_{\ell}$ be a set of probe operators, with expectation values $<G_{\ell}> =tr(G_{\ell}\rho(X))$. The trace defines the Hilbert-Schmidt inner product of operators. The condition for operators (matrices) to be orthonormal is \ba tr(G_{\ell}G_{k}) = \delta_{\ell k} \quad \text{orthonormal matrices}. \label{ortho} \ea  There are $N^{2}-1$ orthonormal $N\times N$ Hermitian operators, not including the identity. When a complete set of probe operators has been measured, the density matrix is tomographically reconstructed from observables as \ba \rho(X) = \sum_{\ell} \, G_{\ell}tr(G_{\ell}\rho) = \sum_{\ell} \, G_{\ell} < G_{\ell}>. \nn \ea For a pure state density matrix, there exists a basis $\{ G_{\ell} \}$ such that only one term appears in the sum over $\ell$. Then $\rho_{pure} =|\psi><\psi|,$ and $|\psi>$ is reconstructed as the eigenvector of $\rho_{pure}$. 

Each orthogonal probe operator measures the corresponding component of the unknown system, and is classified by its transformation properties. For angular distributions the transformations of interest are rotations. 
$\rho(\ell)$ contains tensors transforming like spin-0, spin-1 and spin-2. Each tensor of a given type is orthogonal to the others.

Organizing transformation properties simplifies things significantly. Recall the general form of $\rho(\ell)$, from Eq. \ref{rholep}. The most general form  for $\rho(X)$ that is observable will have the same general expansion, with new parameters: 
  \ba & \text{Probe:} \quad \rho_{ij}(\ell) = {1\over 3}\delta_{ij}  +b \hat \ell \cdot \vec J_{ij} +a U_{ij}(\hat \ell); \quad \text{where} \quad U_{ij}(\hat \ell)= {\delta_{ij} \over 3} -\hat \ell_{i}\hat \ell_{j} =U_{ji}(\ell); \, tr(U(\ell)) =0;  \label{lastline} \\&  \text{System:} \quad  \rho_{ij}(X) = {1\over 3}\delta_{ij}  +{1\over 2}\vec S \cdot \vec J_{ij} +U_{ij}(X) ;  \quad \text{where} \quad U(X)=U^{T}(X); \quad tr(U(X))=0.  \label{nextline} \ea These formulas reiterate Eq. \ref{rholep} while identifying $(J_{k})_{ij} = -\imath \epsilon_{ijk}$ as the generator of the rotation group in the $3\times 3$ representation.\footnote{The real Cartesian basis for $\vec J$ is being used because it is more transparent than the $J_{z} \ra m$ basis that is an alternative. It would have complex parameters.} Upon taking the trace as an inner product, orthogonality selects each term in $\rho(X)$ that matches its counterpart in $\rho(\ell)$. For example $\vec J$ is orthogonal to all the other terms except the same component of $\vec J$: \ba  & {1\over 2} tr(J_{i}J_{k}) =\delta_{ij} ;\nn \\ &\text{hence} \quad {1\over 2} tr( \hat \ell \cdot \vec J  \, \vec S\cdot \vec J ) = \hat \ell \cdot \vec S. \nn \ea Orthogonality makes it trivial to predict which density matrix terms can be measured by probe matrix terms. We call the matching of terms ``the mirror trick.''

We now make several relevant comments about Eq. \ref{lastline} and Eq. \ref{nextline}: \bit \im All density matrices can be written as $1_{N\times N}/N$ to take care of the normalization, plus a traceless Hermitian part. The unit matrix is the spin-0 part and invariant under rotations. The only contribution of the $1$ terms is $tr( 1\times 1)/N^{2} =1/N$. \im  The textbook density matrix {\it spin vector} $\vec S$ consists of those parameters coupled to the angular momentum operator. This is also called the spin-1 contribution. The quantum mechanical average angular momentum of the system is \ba <\vec J> =tr(\rho_{X}\vec J) =\vec S. \nn \ea 
When the coordinates are rotated, the $\vec J$ matrices transform exactly so that $\vec S$ rotates like a vector under proper rotations, and a pseudovector under a change of parity. \im The last term of Eq. \ref{lastline}, the spin-2 part, is real, symmetric and traceless. By the mirror trick it can only communicate with a corresponding spin-2 term in $\rho(X)$ denoted $U_{ij}(X)$, which is real, symmetric and traceless. It can be considered a measure of angular momentum fluctuations: \ba < {1\over 2} \left( J_{i}J_{j}+ J_{j}J_{i}  \right)- {1\over 3}\vec J^{2}\delta_{ij}> = U_{ij}(X). \nn \ea 
A common mistake assumes the quadrupole $U$ should be zero in a pure ``spin state.'' Actually a pure state with $|\vec S|=1$ has a density matrix \ba \rho_{pure, \, ij}(\vec S) ={1\over 2} (\delta_{ij}- \hat S_{i}\hat S_{j})-{ \imath \over 2}\epsilon_{ijk}\hat S_{k}. \label{pure} \ea For example, when $\vec S =\hat z$ the density matrix has one circular polarization eigenstate with eigenvalue unity, and two zero eigenvalues. Pure states exist with $\vec S=0$: They have real eigenvectors corresponding to linear polarization. From the spectral resolution $\rho(X) =\sum_{\a} \, \lambda_{\a} |e_{a}><e_{\a}|$, there is no observable distinction between a density matrix and the occurrence of pure states $|e_{a}>$ with probabilities $\lambda_{\a}$, which are the density matrix eigenvalues. \im As it stands the $U_{ij}$ matrices in Eq. \ref{lastline} and Eq. \ref{nextline} have not been expanded in a complete set of symmetric, orthonormal $3 \times 3$ matrices. Regardless $\rho(X)$ can be fit to data whether or not an expansion is done. The purpose of such work is to complete the classification process to assist with interpreting data. We sketch the steps here. Details are provided in an Appendix. Let $E_{M}$ be a basis of traceless orthonormal matrices where $U(\ell) = \sum_{M} \, tr(U(\ell)E_{M}) E_{M}$. This is the tomographic expansion of the probe. Choose $E_{M}$ so the outputs are normalized real-valued spin-2 spherical harmonics $Y_{M}(\theta, \,  \phi)$. The expansion of the unknown system will be $U(X) =  \sum_{M} \, tr(\rho(X)E_{M}) E_{M} =\sum_{M} \, \rho_{M}(X)E_{M}$. By orthogonality the spin-2 contribution to the angular distribution will be \ba {d N\over d \Omega } \sim tr(\rho(\ell) \rho(X))_{spin-2} \sim \sum_{M} \, \rho_{M}(X) Y_{M}(\theta, \,  \phi). \nn \ea Writing out the terms gives \ba  {dN \over d \Omega}=& {1\over 4 \pi}+\frac{3}{4 \pi}S_{x}\sin\theta \cos\phi+\frac{3}{4 \pi}S_{y}\sin\theta \sin\phi+\frac{3}{4 \pi}S_{z}\cos\theta  \nn \\ &+c\rho_{0}  ({1\over \sqrt{3}} -  \sqrt{3} \cos^2\theta)  - c \rho_{1} \sin(2 \theta)  \cos\phi 
+ c\rho_{2}  \sin^{2}\theta\cos(2 \phi) \nn \\ & +c\rho_{3} \sin^{2}\theta  \sin(2 \phi) - c \rho_{4} \sin(2 \theta)  \sin \phi .\label{res1} \ea The label $X$ has been dropped in $\rho_{M}$ and $c= 3 /(8\sqrt{2} \pi ) $. 
Since $E_{M}$ transform like $Y_{M}$, the coefficients $\rho_{M}$ transform under rotations like spin-2. That means $\rho_{M} \ra R^{(2)}_{MM'}\rho_{M'}(X)$, where $ R^{(2)}_{MM'}$ is a matrix available from textbooks \cite{book}. The traditional $A_{k}$, $\lambda_{k}$ conventions do not use orthogonal functions. Transformations from the traditional conventions to the $\rho_M$ convention are given in an Appendix.

\eit

Note the transformation properties listed are {\it exact}. The systematic and statistical errors of a measurement appear in fitting $\rho(X)$. 

\subsection{Fitting $\rho(X)$, $dN/d\Omega$}

Quantum mechanics requires $\rho(X)$ must be {\it positive}, which means it has positive eigenvalues. Positivity produces subtle non-linear constraints, similar to unitarity. In the $3\times 3$ case the relations are generally cubic polynomials. Positivity is not the same concept as yielding a positive cross section, and generally {\it is a more restrictive} set of relations.\footnote{When $tr(\rho(X))=1$ is maintained, positivity is violated when one or more eigenvalues of $\rho(X)$ exceeds unity, and one or more goes negative. Then for some vector $|e>$ the quadratic form $<e|\rho_{X}|e><0$, which would appear to provide a signal. Yet no such signal might be found in the angular distribution, because $tr(\rho(\ell) \rho(X))>0$ is a much weaker condition. Thus, positivity cannot generally be reduced to bounds on angular distribution coefficients, unless the bounds are so intricately constructed to be equivalent to positivity of the density matrix eigenvalues.} If density matrices are not used it is quite straightforward to fit data yielding a positive cross section while {\it violating positivity}.

Fortunately positivity can be implemented by the Cholesky decomposition of $\rho_X$ \cite{chole}, which is discussed in the Appendix. For the $3\times 3$ case it is: \ba &  \rho(X)(m)=M(m)\cdot M^{\dagger}(m); \nn \\ & 
M(m) = {1 \over \sqrt{ \sum_{k}  m^{2}_k}} \left(
\begin{array}{ccc}
 m_1 & m_4+i m_5 & m_6+i m_7 \\
 0 & m_2 & m_8+i m_9 \\
 0 & 0 & m_3 \\
\end{array}
\right), \label{normed}  \ea where the parameters $-1 \leq m_{\alpha} \leq 1$. 

Event by event $\rho(\ell)$ is an array of numbers, and $\rho(X)(m)$ is an array of parameters. 
The results are combined to make the $J$th instance of $tr(\rho_{J}(\ell)\rho(X)(m))$, where $\rho(X)(m)$ has been parameterized in Eq. \ref{normed}. Fit the $m_{\a}$ parameters to the data set. For example, the log likelihood ${\cal L}$ of the set $J=1...J_{max}$ is \ba {\cal L}(m) =\sum_{J}^{J_{max}} \,  log\left( tr\left(\rho^{(J)}(\ell) \cdot \rho(X)(m) \right)  \right)+ J_{max}log(3/4\pi) . \label{like} \ea Sample code available online\footnote{To help readers appreciate the practical value of these advantages, we constructed standalone analysis code in both ROOT and Mathematica \cite{url}. We expect the code to provide useful cross-checks on code users might write for themselves. \label{link}} carries out these steps, returning parameters $m_{\a}$.
The details of cuts and acceptance appear in fitting the {\it numbers} $m_{\a}$ using {\it numbers} for the lepton matrix $\rho(lep)$ (not angles, nor trigonometric functions.) In one example with simulated $Z$-boson data we found \ba \rho_{fit}(X) = \left(
\begin{array}{ccc}
 0.5574 & 0.01399-0.07144 i &
   -0.004026+0.013487 i \\
 0.01399+0.07144 i & 0.4422 & 0.003138-0.002670
   i \\
 -0.004026-0.013487 i & 0.003138+0.002670 i &
   0.0004268 \\
\end{array}
\right) \nn \ea Using the Standard Model parameters for $\rho(\ell)$, Eq. \ref{hatell} and Eq. \ref{rholep}, the trace yields \ba & {d N \over d \Omega_{fit}}\sim tr(\rho(\ell)\rho(X))=0.5000+ 0.0007739 \sin (\phi ) \sin (\theta )+0.3090 \cos
   ^2(\phi ) \cos ^2(\theta ) \nnn +0.1904 \sin ^2(\phi )
   \cos^2(\theta )+... \nn \ea where ... indicates several terms there is no need to write out. Integrated over $\phi$, this expression becomes \ba {dN_{fit} \over d \cos\theta } ={3\over 4 \pi}\left( 1.57 + 0.137 \cos \theta + 1.56 \cos^{2} \theta \right). \nn \ea A $1+cos^{2}\theta$ distribution is the leading order Drell-Yan prediction for virtual spin-1 boson annihilation, while $ 0.137 \, cos\theta$ represents a charge asymmetry.  

It is trivial to go from $tr(\rho(\ell)\rho(X))$ to a conventional parameterization of an angular distribution by taking inner products of orthogonal functions. It is also easy to expand $\rho(X)$ in a basis of orthonormal matrices with the same results. Note these steps are {\it exact}, and much different from fitting data to trigonometric functions in some convention, which tends to yield multiple solutions, along with violations of positivity, which can introduce pathological convention-dependence. Perhaps struggles with convention-dependence of quarkonium data \cite{Faccioli:2010ji, Faccioli:2010kd} are related to this. It would be interesting to investigate.

\subsection{Summary of quantum tomography procedure}
\label{sec:summary}

To analyze data for each event labeled $J$: \bit \im
$\text{Compute} \; Q_{J} =k_{J}+k_{J}'; \; \; \; \; \ell_{J} =k_{J}-k_{J}'; \; \; \; \; (X_{J}^{\mu}, \, Y_{J}^{\mu}, \, Z_{J}^{\mu}) ; \smallskip \\ \smallskip
 \vec \ell_{XYZ, J} = ( X_{J}\cdot \ell_{J}, \, Y_{J}\cdot \ell_{J} , \, Z_{J}\cdot \ell_{J});\\
\hat \ell_{J} =\ell_{XYZ, \, J}  /\sqrt{-\ell_{XYZ, \, J}\cdot \ell_{XYZ, \, J}}.$

 \im  Make the lepton density matrix. For $Z$ bosons in the Standard Model it is \ba \rho_{ij}(\ell) = {1 \over 2}(\delta_{ij} - \hat \ell_{i} \hat \ell_{j}) - 0.22\, \imath  \epsilon_{ijk} \hat \ell_{k}. \ea  
 \im The results are combined to make the $J$th instance of $tr(\rho_{J}(\ell)\rho(X)(m))$, where $\rho(X)(m)$ has been parameterized in Eq. \ref{normed}. Fit the $m_{\a}$ parameters to the data set. For example, the log likelihood ${\cal L}$ of the set $J=1...J_{max}$ is \ba {\cal L}(m) =\sum_{J}^{J_{max}} \,  log\left( tr\left(\rho^{(J)}(\ell) \cdot \rho(X)(m) \right)  \right)+ J_{max}log(3/4\pi) . \label{like} \ea Sample code available online (see footnote \ref{link}) carries out these steps, returning parameters $m_{\a}$.   \eit

\subsection{Comments}
1. The possible symmetries of $\rho(\ell)$ enter here. Suppose $c_{A}=0$. Then $\rho(\ell)$ is even under parity, real and symmetric. The imaginary antisymmetric elements of $\rho(X)$ are orthogonal, and contribute nothing to the angular distribution. When known in advance, the redundant parameters of $\rho(X)$ can be set to zero while making the fit. (That does not mean unmeasured parameters can be forgotten when dealing with positivity.)
In general a fitting routine will either report a degeneracy for redundant parameters, or converge to values generated by round-off errors. Degeneracy will always be detected in the Hessian matrix computed to evaluate uncertainties. 

2. The normalization condition $ \sum_{k} \, m^{2} (k)=1$ can be postponed by removing $1/\sqrt{m_{\a}^{2}}$ from Eq. \ref{normed}, and subtracting $J_{max}log( \sum_{k}^{J_{max}} \, m^{2} (k))$ from the log-likelihood (Eq.\ref{like}). When that is done the fitted density matrix will not be automatically normalized, due to the symmetry $\rho(X) \ra \lambda \rho(X)$ of the modified likelihood. The density matrix becomes normalized by dividing by its trace.   Incorporating such tricks improved the speed of the code available online (see footnote \ref{link}) by a factor of about 100.

3. Algorithms are said to compute a ``unique'' Cholesky decomposition, which would seem to predict $m_{\a}$ given $\rho(X)$. The algorithms choose certain signs of $m_{\a}$ by a convention making the diagonals of $M$ positive. However that is not quite enough to assure a numerical fit finds a unique solution. 

The fundamental issue is that $MM^{\dagger}=\rho(X)$ is solved by $M= \sqrt{\rho(X)}$, and the square root is not unique. There are $2^{N}$ arbitrary sign choices possible among $N$ eigenvalues of $\sqrt{\rho(X)}$. Forcing the diagonals of $M$ to be positive reduces the possibilities greatly, and an algorithm exists to force a unique, canonical form of $m_{\a}$ in a data fitting routine. We did not make use of such a routine, since fitting $\rho(X)$ is the objective. Depending upon the data fitting method, increasing the number of ways for $M(m_{\a})$ to make a fit sometimes makes convergence faster.  

4. Let $<>_{exp}$ stand for the expectation value of a quantity in the experimental distribution of events. By symmetry $ <\hat \ell>_{exp}$ and $< \hat \ell_{i} \hat \ell_{j} >_{exp}$ are vector and tensor estimators, respectively, which must depend on the vector and tensor parameters $\vec S$, $U_{ij}(X)$ in the underlying density matrix. A calculation finds \ba &  <\hat \ell>_{exp} = {1 \over J_{max}}\sum_{J} \hat \ell_{J} = - {1 \over 4}\vec S; \nn \\ & < \hat \ell_{i} \hat \ell_{j} >_{exp}={1 \over J_{max}}\sum_{J} \hat \ell_{Ji}\hat \ell_{Ji} = {1\over 3}\delta_{ij} - {1\over 5} Re[U_{ij}]. \nn \ea An estimate of $\rho(X)$ not needing a parameter search then exists directly from data. However positivity of $\rho(X)$ is more demanding, and not automatically maintained by such estimates.

\section{Results}
\subsection{Analysis Bonuses of the Quantum Tomography Procedure}

\subsubsection{Convex Optimization}
The issue of multiple solutions for $\rho(X)$ is different. Multiple minima of $\chi^{2}$ statistics affects fits to cross sections parameterized by trigonometric functions. However, quantum tomography using maximum likelihood happens to be a problem of {\it convex optimization}. In brief, when $\rho$ is positive then $<e|\rho|e>$ is a positive convex function of $|e>$. Then $tr(\rho(\ell) \rho(X))$ is convex, being equivalent to a positively weighted sum of such terms. The logarithm is a concave function, leading to a convex optimization problem. That means that {\it when $\rho(X)$ is a local maximum of likelihood it is the global maximum.} Exceptions can only come from degeneracies due to symmetry or an inadequate number of data points \cite{DBLP:journals/mp/BurerM05}. Convex optimization is  
important because without such a property the evaluation of high-dimensional fits by trial and error can be exponentially difficult.

\subsubsection{Discrete Transformation Properties} 

\begin{small} 

\begin{table*}[ht]
\centering

$\begin{array}{|lll || c | c | c | c | c |}
\hline
term & origin & dN/d\Omega & C_{\ell}  & P & T & C_{\ell}P &PT \\ 
\hline 
 \cdot & \ell & \cdot & -  &   - & -   & +  & +  \\  
 \cdot &  X & \cdot  & \cdot  &   -& -   &+   & + \\   
 \cdot  &  Y &\cdot  &  \cdot  &  + & +  &+    & +  \\ 
 \cdot  &   Z &  \cdot  &  \cdot    & -  & -   & +   & + \\ 
S_{x}&   X\ell &  \sin\theta \cos \phi  &-  & +  & +  & -  & +\\ 
S_{y}  &  Y\ell  &  \sin\theta \sin \phi  & - & -  & -   & + & + \\ 
S_{z}  &   Z\ell  &  \cos\theta    &  - & +  &+   & -  & + \\ 
\rho_{2} &   XX \ell \ell & \sin^{2}\theta\cos 2\phi    &    + &  + & +   &  +  &+ \\ 
\rho_{3}   &    XY \ell \ell   &  \sin^{2}\theta\sin2\phi      &  +  &  - &  - &  +  & + \\ 
\rho_{1}   &     XZ \ell \ell  & \sin 2\theta \cos\phi        &    + &  + & +   &  +  & + \\ 
\rho_{4}     &   YZ \ell \ell  &  sin 2\theta\sin\phi      &  +  &  - &  - &  +  & + \\ 
\rho_{0}     &    ZZ\ell \ell   & 1/\sqrt{3} -\sqrt{3} \cos^{2}\theta     &    + &  + & +   &  +  & + \\  \hline 
  \end{array}$

  \caption{ \small Terms in the angular distribution with their properties under discrete transformations $C_{\ell}$, $P$, and $T$. Here $\ell$ stands for $\hat \ell$, $ X\ell $ stands for $\hat X \cdot \hat \ell=-X_{\mu} \ell^{\mu}$, and so on with scalar normalization factors removed. $T$-odd scattering observables from the imaginary parts of amplitudes generally exist without violating fundamental $T$ symmetry. See the text for more explanation. }\label{tab:symmetries}
\end{table*}   \end{small} 

Table \ref{tab:symmetries} lists discrete transformation properties of all terms under parity $P$, time reversal $T$, and
lepton charge conjugation $C_{\ell}$. If leptons have different flavors (as in like or unlike sign $e \mu$) the $C_{\ell}$ operation swaps the particle defining $\hat \ell$.  

When coordinates $XYZ$ are defined the direction of $\hat Y =\hat Z\times \hat X$ is even under time reversal and parity, which is exactly the opposite of $X$ and $Z$. Then $\vec S \cdot \hat Y$ is $T$-odd, contributing the $\sin \theta \sin\phi$ term.\footnote{In a forthcoming study \cite{INT} of inclusive lepton pair production near the $Z$ pole, we find interesting, new features in the $S_y$ data of Ref.  \cite{Aad:2016izn}.} The $XY$ and $ZY$ matrix elements of $\rho(X)$ are also odd under $T$, contributing the terms shown. $T$-odd terms come from imaginary parts of amplitudes, which are generated by loop corrections in perturbative QCD.

Notice that every term in the lepton density matrix (Eq. \ref{rholep}) is automatically symmetric under $C_{\ell}P$. This is a kinematic fact of the lepton pair probe which does not originate in the Standard Model. As a result the $C_{\ell}P$ transformations of the angular distribution depend on the coupling to the unknown system. If overall $CP$ symmetry exists the target density matrix will have $CP$ odd terms where $C_{\ell}P$ odd terms are found. In the Standard Model these $\cos\theta$ and $\sin \theta \cos \phi$ terms correspond to charge asymmetries of leptons correlated with charge asymmetries of the system, namely the beam quark and anti-quark distributions.

While weak $CP$ violation is a mainstream topic, $P$ and $CP$ symmetry of the strong interactions at high energies has not been tested \cite{mihailo}.  The gauge sector of $QCD$ is {\it kinematically} $CP$ symmetric, because the non-Abelian $tr(\vec E\cdot \vec B)$ term is a pure divergence.\footnote{Non-perturbative strong $CP$ violation in $QCD$ by a surface term has been proposed. Tests have been dominated by the neutron dipole moment, while calculations of non-perturbative effects are problematic.}. Higher order terms in a gauge-covariant derivative expansion are expected to exist, and can violate $CP$ symmetry \cite{mihailo}.

However, measuring violation of $CP$ or fundamental $T$ symmetry in scattering experiments is invariably frustrated by the experimental impossibility of preparing time-reversed counterparts. Some ingenuity is needed to devise a signal. It appears that any signal will involve four independent 4-momenta $p_{J}$ and a quantity of the form $\Omega_{4}= \epsilon_{\a \b\lambda \sigma } p_{1}^{\a}p_{2}^{\b}p_{3}^{\lambda}p_{4}^{\sigma}$. For example a term going like $\ell \cdot Y \sim \epsilon_{\a \b\lambda \sigma}\ell_{\a}Q_{\b}P_{A \lambda}P_{B \sigma}$ might possibly originate in fundamental $T$ symmetry violation, and be mistaken for perturbative loop effects. A more creative road to finding $CP$ violation involves two pairs with sum and difference vectors $Q, \, \ell; \, Q', \, \ell'$, and the scalar $\epsilon_{\a \b\lambda \sigma}\ell_{\a}Q_{\b}Q_{\lambda}' \ell'_{ \sigma}$, which is even under $C$ and odd under $P$. The pairs need not be leptons (although ``double Drell Yan'' has long been discussed) but might be (say) $\mu^{+}\mu^{-} \pi^{+}\pi^{-}$. It would be interesting to explore further what a tomographic approach to such observables might uncover.

\subsubsection{Density Matrix Invariants}
\label{sec:Invariants} 

We mentioned that scattering planes, trig functions, boosts and rotations could be avoided, and the examples show how. Once a frame convention is defined the lepton ``coordinates'' $( X_{J}\cdot \ell_{J}, \, Y_{J}\cdot \ell_{J} , \, Z_{J}\cdot \ell_{J})$ are actually Lorentz scalars. However they depend on the convention for $XYZ$, which is arbitrary. At least four different conventions compete for attention. Moreover, once a frame is chosen, at least two naming schemes (the ``$A_{k}$'' and ``$\lambda_{k}$'' schemes) exist to describe the angular distribution in terms of trigonometric polynomials..

Well-constructed invariants can reduce the confusion associated with convention-dependent quantities \cite{Palestini:2010xu, Shao:2012fs, INT, Ma:2017hfg}. Since $\vec S$ transforms like a vector its magnitude-squared $\vec S^{2}$ is rotationally invariant. The spin-1 part of $\rho(X)$ does not mix with the real symmetric part under rotations. Since it is traceless, the real symmetric (spin-2) part has two independent eigenvalues, which are rotationally invariant.\footnote{Work by Faccioli and collaborators \cite{Faccioli:2010ps, Faccioli:2010ej} attempted to construct invariants by inspecting the transformation properties of ratios of sums of angular distribution coefficients upon making rotation about the conventional $Y$ axis. The method cannot identify a true invariant unless $Y$ happens to be an eigenvector of the matrix. By the same method the group also identified $\vec S^{2}$ as a ``parity violating invariant,''  while $\vec S^{2}$ is actually even under parity. Parity violation is not required to measure $\vec S$ with polarized beams.} Finally the dot-products of three eigenvectors $\hat e_{J}$ of the spin-2 part with $\vec S$ are rotationally invariant. Then $(\hat e_{j}\cdot \vec S)^{2}$ are three invariants not depending on the sign of eigenvectors. That suggests six possible invariants, but $\sum_{j} \, (\hat e_{j}\cdot \vec S)^{2}=\vec S^{2}$ makes the $\vec S$ invariants dependent, leaving five independent rotational invariants. That is consistent with counting 8 real parameters in a $3\times 3$ Hermitian matrix, subject to 3 free parameters of the rotation group, leaving 8-3=5 rotational invariants. The same counting for unitary transformations would leave only the two independent eigenvalues of the matrix.

\begin{figure}[htp]
\begin{center}
\includegraphics[width=4in]{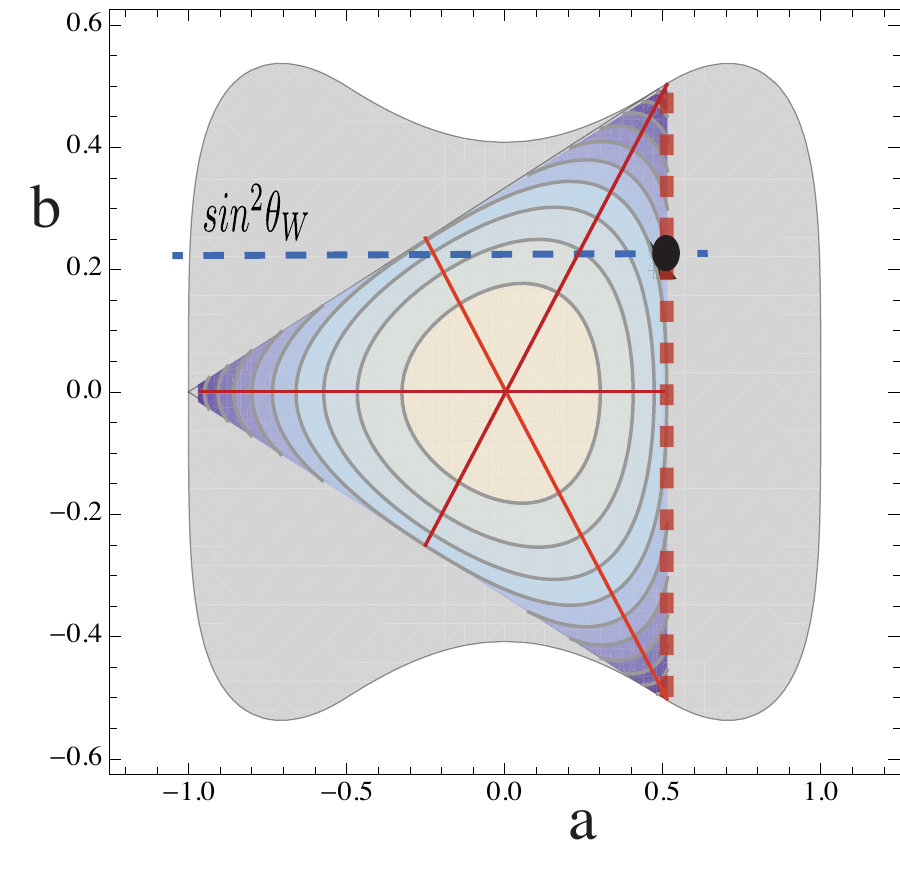}
\caption{ Contours of constant entropy ${\cal S}$ of the lepton density matrix $\rho(\ell)$ (Eq. \ref{rholep}) in the plane of parameters $(a, \, b)$. Contours are separated by 1/10 unit with ${\cal S}=0$ at the central intersection. The horizontal dashed line shows the lowest order Standard Model prediction $ b=sin^{2}\theta_{W}$. Annihilation with on-shell helicity conservation is indicated by the vertical dashed line $a=1/2$. The left corner of the triangle is a pure state with longitudinal polarization, while the two right corners are pure states of circular polarization. The interior lines represent matrices with maximal symmetry, where two eigenvalues are equal. They cross at the unpolarized limit. The curved gray region represents the much less restrictive constraints of a positive distribution using Eq. \ref{dist1} and lepton universality. }
\label{fig:EntyContours}
\end{center}
\end{figure}

Any function of invariants is invariant. The combinations below have useful physical interpretations: \bit \im  The {\it degree of polarization $d$} is a standard measure of the deviation from the unpolarized case. It comes from the sum of the squares of the eigenvalues of $\rho$ minus 1/3, normalized to the maximum possible: \ba & d = \sqrt{ (3 tr(\rho_X^{2})-1)/2} ,\nn \\ & \qquad \text{where} \qquad 0 \leq d \leq 1. \nn \ea When $d=0$ the system is unpolarized, and when $d=1$ the system is a pure state.  

\im The {\it entanglement entropy ${\cal S}$} is the quantum mechanical measure of order. The formula is \ba & {\cal S}=-tr(\rho_X \, log(\rho_X)) \nn  \ea  In terms of eigenvalues $\rho_{\a}$, ${\cal S} = -\sum_{\a } \, \rho_{\a} log(\rho_{\a})$.
When $\rho \ra 1_{N\times N}/N$ the system is unpolarized, and ${\cal S}=log(N)$. That is the maximum possible entropy, and minimum possible information. When ${\cal S}=0$ the entropy is the minimum possible, providing the maximum possible information, and the system is a pure state. 

It is instructive to interpret $e^{{\cal S}}$ as the ``effective dimension'' of the system. For example the eigenvalues $(1/2 +b, \, 1/2-b, \, 0)$ occur in the density matrix of on-shell fermion annihilation with helicity conservation. One zero-eigenvalue describes an elliptical disk-shaped object. The entropy ranges from ${\cal S}=0$, ($e^{{\cal S}}=1$ for $b=1/2$, a one dimensional stick shape) to ${\cal S}=log(2)$, ($e^{{\cal S}}=2$ for a disk-shaped object with maximum symmetry.) As expected, an unpolarized 3-dimensional system has three equal eigenvalues, is shaped like a sphere, and $e^{{\cal S}} \ra 3.$ 

\begin{figure}[htp]
\begin{center}
\includegraphics[width=4in]{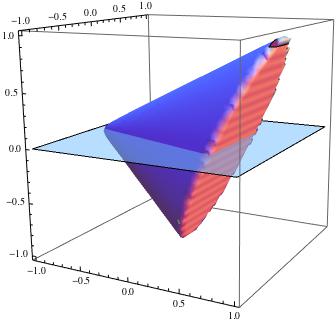}
\caption{Boundary of the positivity region of a density matrix depending on three parameters $a, \, b, \, c$ described in the text. The two-dimensional region cut by the plane $c=0$ corresponds to Figure \ref{fig:EntyContours}.  }
\label{fig:PlaneCut}
\end{center}
\end{figure}

Figure \ref{fig:EntyContours} shows the entropy of the lepton density matrix $\rho(\ell)$ (Eq. \ref{rholep}) in the plane of parameters $(a, \, b)$. The matrix eigenvalues are\footnote{It can be shown that the eigenvalues $\lambda_{k} =1/3+( 2d/3) \cos(\theta_{k} )$, where $d$ is the degree of polarization and $\theta_{k} = \cos^{-1}(det( (3\rho(X)-1_{3\times 3})/d)/2+ 2 \pi k/3)$.} $(1/3 - 2 a/3,  \, 1/3 + a/3 - b, \, 1/3 + a/3 + b)$. The triangular boundaries are the positivity bounds on these parameters, outside of which the entropy has an imaginary part. The corners of the triangle are pure states. The left corner represents a purely longitudinal polarization, $\rho_{L}=|L><L|$ where $|L>=(0, \, 0, \, 1)$ in a coordinate system where $\hat \ell=\hat z$. The two right corners are purely circular polarizations, $\rho_{\pm}=|\epsilon_{\pm}><\epsilon_{\pm}|$, where in the same coordinates $ \epsilon_{\pm} =(1, \, \pm \imath, \, 0)/\sqrt{2}.$ The interior lines $a=\pm b, \, b=0$ represent maximal symmetry matrices having two equal eigenvalues. The figure also indicates the constraints of a positive distribution for the example of Eq. \ref{dist1} assuming lepton universality. The values of $a$ and $b$ are actually unrestricted in all directions, so long as they lie within the bounding curves.

The Standard Model leptons from lowest order $s$-channel $Z$ production have $a=1/2, \, b=\sin^{2}\theta_{W}$, which is shown in Fig. \ref{fig:EntyContours} as a dot. The edge $a=1/2$ corresponds to on-shell helicity conservation, with eigenvalues $0,  \, 1/2 - \sin^{2}\theta_{W}, \, 1/2 + \sin^{2}\theta_{W}$. The $a, \, b$ parameters of leptons from a different production process, or subject to radiative corrections, must still lie inside the triangle. Maximal symmetry with eigenvalues (1/2, 1/2, 0) occurs where the line of $b=\sin^{2}\theta_{W}$ just touches the $b=-a$ line, which happens at $\sin^{2}\theta_{W}=1/4$. That is not far from the Standard Model value, which is very interesting. Since no established theory predicts $\sin^{2}\theta_{W}$ one cannot rule out a deeper connection. 

It is tempting but incorrect to assume the bounds discussed would apply to the same terms of a more general density matrix. For example, add $-c \hat n_{i} \hat n_{j}$ to the expression in Eq. \ref{rholep}, where $\hat n \cdot \hat \ell=0$ and update the normalization condition. The resulting positivity region of $a, \, b, \, c$ is shown in Figure \ref{fig:PlaneCut}, which also shows the plane $c=0$ equivalent to Figure \ref{fig:EntyContours}. At the extrema $c=\pm 1$ the region of consistent $(a, \, b)$ parameters shrinks to single points.

The matrix for $\rho(X)$ computed earlier is an example where all terms in any standard convention happen to occur. By inspection this system (mostly quark-antiquark annihilation) is superficially much like the lepton one. The entropy of is 0.68 and $e^{{\cal S}}=1.96$, and one eigenvalue is close to zero. Of course there is much more information in the other parameters, the orientation of eigenvectors, $\vec S$, and its magnitude.\eit

\section{Discussion} 

The quantum tomography procedure offers {\it at least seven} significant advantages over standard methods of analyzing the angular correlations of inclusive reactions:

\begin{itemize}
\item{\it Simplicity and Efficiency.} Tomography exploits a structured order of analysis. By construction, unobservable elements never appear.  

\item{\it Covariance.} Physical quantities are expressed covariantly every step of the way. That is not always the case with quantities like angular distributions.

\item{\it Complete polarization information.} The unknown density matrix $\rho(X)$ contains all possible information, ready for classification under symmetry groups.

\item{\it Model-independence.} No theoretical planning, nor processing, nor assumptions are made about the unknown state. The process of defining general structure functions has been completely bypassed. It is not even necessary to assume anything about the spin of $s$- or $t-$channel intermediates. The {\it observable} target structures is always a mirror of the probe structure. The ``mirror trick'' is universal as described in Section \ref{sec:mirrror}. 

\item{\it Manifest positivity.} A pattern of misconceptions in the literature misidentifies positivity as being equivalent to positive cross sections. It is not difficult to fit data to an angular distribution and violate positivity. In fact, {\it an angular distribution expressed in terms of expansion coefficients actually lacks the quantum mechanical information to enforce positivity.}  

\item{\it Convex optimization.} The positive character of the density matrix leads to convex optimization procedures to fit experimental data. This provides a powerful analysis tool that ensures convergence..

\item{ \it Frame independence.} Once the unknown density matrix has been reconstructed, rotationally invariant quantities can be made by straightforward methods. This is illustrated in Section \ref{sec:Invariants}, which includes a discussion of the entanglement entropy.

\end{itemize}

Quantum tomography has already yielded significant results. Our tomographic analysis \cite{INT} of a recent ATLAS study of Drell-Yan lepton pairs with invariant mass near the $Z$ pole \cite{Aad:2016izn} discovered surprising features in the density matrix eigenvalues and entanglement entropy. By way of advertising, we have also gained insight into the mysterious Lam-Tung relation \cite{LamTung78,*LamTung80}, including why it holds at NLO but fails at NNLO. These topics will be presented in separate papers.

\begin{acknowledgments}
The authors thank the organizers of the INT17-65W workshop "Probing QCD in Photon-Nucleus Interactions at RHIC and LHC: the Path to EIC" for the opportunity to present this work. We also thank workshop participants for useful comments.
\end{acknowledgments}

\section*{References}
\bibliography{tomogbiblio}

\section{Appendix}
\subsection{Cross section in terms of density matrices}
The density matrix approach stands on its own as an efficient tool kof quantum mechanics. However we see the need to relate it to more traditional scattering formalism.

Consider the cross section for the inclusive production of two final-state particles of spin $s$, $s'$ from particle intermediates:
$$d\sigma \sim \sum_{s,s'}|\sum_J {\cal M}(\chi_J \ra f_{s,s'})|^2 \cdot d \Pi_{LIPS},$$
where $\sum_J {\cal M}(\chi_J \ra f) = \sum_J  <f|T|\chi_J> \delta^4(\sum p_f -\sum p_i),$ $T$ is a transfer matrix, and $d\Pi_{LIPS}$ is the Lorentz-invariant phase space.

Then,
\ba d\sigma \sim \sum_{s,s'} [(\sum_J <f_{s,s'}|T|\chi_J>)\cdot ( \sum_K <\chi_K|T^{\dag}|f_{s,s'}>)]\cdot d\Pi_{LIPS}, \label{dsigdef}\ea
$$=tr[(\sum_{s,s'}T^{\dag}|f_{s,s'}><f_{s,s'}|T)\cdot (\sum_{J,K}|\chi_{J}><\chi_{K}|)]\cdot d\Pi_{LIPS},$$
where $\sum_{J,K} |\chi_J><\chi_K|$ accounts for any interference between intermediate states. We identify the quantity in the first set of parentheses on the last line of Eq. \ref{dsigdef} with the probe density matrix $\rho_{probe}$ and the quantity in the second set of parentheses with the density matrix for the unknown particle intermediates, $\rho_{X}$. The $T$, $T^{\dag}$ in $\rho_{probe}$ ensure the overlap of $\rho_X$ with $\rho_{probe}$, inside the trace, is taken at 'equal times'.

Rewriting $d\sigma$ in terms of the density matrices, we find,
$$d\sigma \sim tr(\rho_{probe}\cdot \rho_X) \cdot d\Pi_{LIPS},$$
where $d\Pi_{LIPS} \sim d\Omega \cdot d^4q$ and $q$ is the sum of the final-state pair momentum. 
Then, for given pair momentum $q$, the angular distribution is,
$$\frac{d\sigma}{d\Omega}(\theta, \phi | q) \sim tr(\rho_{probe} \cdot \rho_X),$$
where the proportionality suppresses an overall normalization. The conditional probability {\it given} $q$ captures the event-by-event character of angular correlations. The explicit $q$ dependence might suggest we assumed an $s$-channel boson intermediate state, but we have not.

If we know $d\sigma/d\Omega$ and $\rho_{probe}$ for a given event, we can reconstruct $\rho_X$ for that event. This is the essence of the quantum tomography procedure. The probe is what is known, and it determines what can be discovered.

\subsection{Probe matrix}
\begin{figure}[ht]
\begin{center}
\includegraphics[width=4in,height=2in]{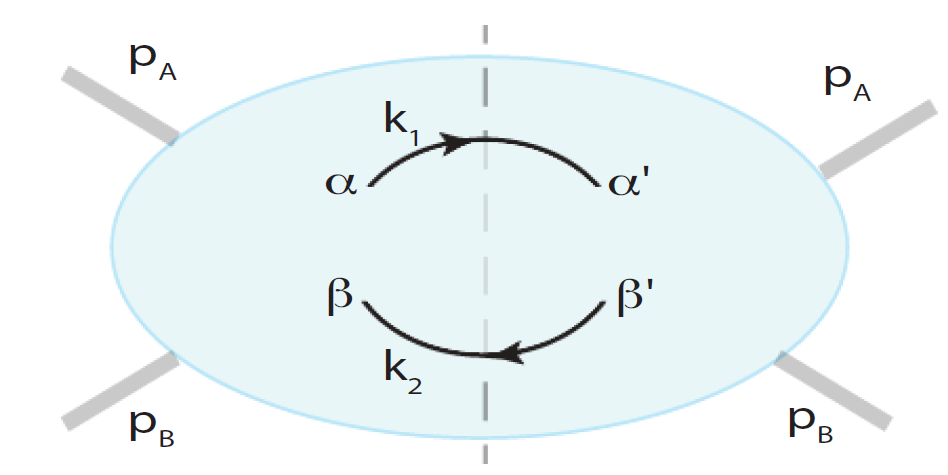}
\caption{ The lepton pair density matrix $\rho^{\a \a}_{\b\b'}(lep)$, in black, coupled to the colliding system density matrix $\rho(X)$. The matrix labels on legs are diagonal in momenta $k_{1}$, $k_{2}$. Off-diagonal polarization (Dirac) indices are explicitly shown. The Feynman rules are the same as for ordinary diagrams.  }
\label{fig:Probe1}
\end{center}
\end{figure}

Figure \ref{fig:Probe1} shows the diagram for the simplest lepton-pair probe. It is completely non-specific about the process  creating a lepton with momentum $k_{1}$ and Dirac density matrix polarization $\a \a'$, and an anti-lepton with  momentum $k_{2}$ and polarization $\b \b.'$ From the Feynman rules \ba \rho(lep)_{\a \a'}^{\b\b'}\sim (\slasha k_{1})_{\a \a'}(\slasha k_{2})_{\b \b'}. \label{probehere} \ea The symbol $\sim$ indicates the high energy limit and ignoring a trivial overall normalization. 

Continuing, $\rho(X)$ is something of vast complexity, which can only couple to $\rho(lep)$ via the indices shown. 
The Dirac structure of $\rho(X)$ can be expanded over several complete sets. However the relevant (observable) part of $\rho(X)_{\a \a'}^{\b\b'}$ will be its projection {\it onto the subspace coupled to this particular probe}, just as the general analysis prescribed. It is ideal to classify only what will be observed. That sector is predetermined by the very limited Dirac structure of the probe. Thus \ba & d\sigma \sim k_{1}^{\mu}k_{2}^{\nu} \rho_{\mu \nu}(X), \nn \\ & \qquad \text{where} \quad \rho_{\mu \nu}(X)= tr\left (( \gamma_{\mu} \otimes \gamma_{\nu}) \,  \rho(X) \right). \nn \ea With this probe massless fermion pairs produce no combinations of $k_{1}^{\mu}k_{1}^{\nu}$ or $k_{2}^{\mu}k_{2}^{\nu}$ or anything else. The operations show how the $tr$ symbol comes to be used repeatedly with different meanings implied by the context. 

We now use Hermiticity, which makes two predictions: \ba \rho(lep) = \rho^{\mu \nu}(lep)\gamma_{\mu}\gamma_{\mu}; \qquad  \rho^{\mu \nu}(lep)=  \rho^{\nu \mu}(lep)^{*}, \nn \ea plus the same relation for $\rho(X)$. All $k_{1}^{\mu}k_{2}^{\nu} $ factors must be strictly bilinear, and occur in a real symmetric plus $\imath \times$ antisymmetric combinations. The most general possibility is \ba  & \rho^{\mu \nu}(lep)= \a k_{1}\cdot k_{2}\eta^{\mu \nu} +\b(k_{1}^{\mu}k_{2}^{\nu} +k_{1}^{\nu}k_{2}^{\mu}) +\imath \gamma (k_{1}^{\mu}k_{2}^{\nu} -k_{1}^{\nu}k_{2}^{\mu})+\imath \delta\epsilon^{\mu \nu \a \b}k_{1 \a}k_{2\b} , \label{kkforma} \ea where $\eta_{\mu \nu}$ is the Minkowski metric. By algebra \ba  \rho^{\mu \nu}(lep)= {\a \over 2}q^{2}\eta^{\mu \nu} + {\b\over 2} q^{\mu}q^{\nu}- {\b\over 2} \ell^{\mu} \ell^{\nu} -\imath {\gamma\over 2}(q^{\mu}\ell^{\nu} -q^{\nu}\ell^{\mu} )- \imath {\delta\over 2} \epsilon^{\mu \nu \a \b}q_{\a}\ell_{\b}. \label{qqforma} \ea 

Event by event there exists a preferred, oriented pair rest frame where $q^{\mu} =(q, \, \vec 0)$ and $\ell^{\mu}=(0, \, q\hat \ell)$. In that frame Eq. \ref{qqforma} predicts a normalized $3\times 3$ tensor of spatial components which is \ba \rho^{jk}(lep)=&{1\over 3}\delta^{{jk}} + a\,J_{p}^{jk}\ell^{p} - b\, U^{jk};  \label{rholep2} \\ & U^{jk}(\hat \ell)= {\hat \ell^{j}\hat \ell^{k}-{\delta^{jk} \over 3}}. \label{udef} \ea Here $a$ and $b$ are real, and $J_{p}$ are the spin-1 rotation generators in Cartesian coordinates: \ba  J_{p}^{jk}= -\imath \epsilon_{pjk}. \nn \ea  Equation \ref{rholep2} has been decomposed into tensors transforming under rotations like spin-0, spin-1 and spin-2. The expansion above is {\it kinematic} and not a consequence of any special theory. 

The approach has the virtue of maintaining strict control of how outputs depend on assumptions. We have made few assumptions, yet we have a result, which is that $\rho^{jk}(lep)$ only depends on two {\it scalar}  $a$, $b$. The only scalar available from $\rho(lep; \, \ell, \, q)$ is $q^{2}$, hence $a=a(q^{2})$, $b=b(q^{2})$. The enormous body of field theory and the Standard Model only predicts only two parameters of Eq. \ref{rholep2}. {\it If and when} the lepton pair originates from an intermediate boson with vertex $c_{V}\gamma^{\mu}+ c_{A}\gamma_{\mu}\gamma_{5}$, then \ba  a=c_{A}c_{V}; \qquad b=1/2. \nn \ea The general possibility these parameters might be functions of $q^{2}$, namely vertex form factors, has emerged on its own. Notice that in the rest frame oriented naturally along the lepton momentum $\rho(lep)$ is not diagonal. The diagonal elements are interpretable as probabilities, even {\it classical} probabilities. The off-diagonal elements convey the information about entanglement.

\subsection{Positivity}

There is a positivity issue in fitting angular distribution data. Represent $\rho =MM^{\dagger}$, and then $\rho >0$. Any $M=HU$, where $H=H^{\dagger}$ and $UU^{\dagger}=1$. We can make $M$ self-adjoint since $U$ cancels out. 
To parameterize $N \times N$ matrices $M$ use $SU(N)$ representations $G_{a}$, normalized to $G_{a}G_{b} =(1/2)\delta_{a b}$. For $N=3$ those are the Gell-Mann matrices. We define parameters with \ba M = m_{0} 1_{3\times 3}/\sqrt{3}+\sqrt{2} m_{a}G_{a}. \label{meq} \ea 

Compute \ba & MM^{\dagger} = m_{0}^{2}/3+2 \sqrt{{2 \over 3}}m_{0}m_{a }G_{a} + 2 m_{a} m_{b}G_{a}G_{b}. \nn \ea The symmetric product is \ba G_{a}G_{b}+G_{b}G_{a} = \delta_{a b}/3+ f_{a bg} G_{g}. \nn \ea Check the trace of both sides for the normalization of $\delta_{a b}$. Everything else must be traceless and spanned by $G_{g}$. Then 
\ba MM^{\dagger} &={1\over 3}(m_{0}^{2} +\sum_{a}m_{a}^{2})+ 
2  \sqrt{{2 \over 3}}m_{0}m_{g }G_{g} + f_{abg}m_{a}m_{b}G_{g}. \nn \ea The normalization $tr(\rho)=1$ needs $\sum_{\mu =0}^{9} \, m_{\mu} m_{\mu}=1$. This requires each $0<m_{\mu}^{2}<1$, while it is more restrictive. 

There is a degeneracy issue in the nonlinear relation of $m_{\mu}$ to a straight expansion $\rho  =1/3+c_{g}G_{g}$, \ba c_{g} =2  \sqrt{{2 \over 3}}\sqrt{1-m\cdot m} \, m_{g } +f_{a bg}m_{a}m_{b} \ea Notice \ba [\rho, M] =  [ M^{2},  \, M] =0. \nn \ea Then $M$ and $\rho$ have the same eigenvectors. The eigenvalues of $\rho$ are those of $M$, squared. For $N$ eigenvalues of $\rho$ there are $2^{N}$ possible $M$'s. If $\rho$ is positive definite, however, there is only one $M$ with strictly positive diagonal entries. In that sense, the $M$ satisfying $\rho = M M^{\dag}$ can be said to be unique.

The positivity problem is often solved with the Cholesky decomposition: $\rho=LL^{\dagger}$, where $L$ is a lower-triangular matrix with real entries on the diagonal. $L$ is related to $M$ by a similarity transform. There are $N+ 2 N(N-1)/2=N^{2}$ free real parameters in a lower-triangular matrix with real diagonals, which is just right for Hermiticity. As before, the condition $tr(\rho)=1$ requires $\sum \, m_{\mu} m_{\mu}=1$. The Cholesky decomposition is unique, in the sense above, when $\rho$ is positive definite.

\subsection{Collected conventions}
As a consequence of consistent definitions, our $\rho_{M}$ and $Y_{M}$ transform under rotations like real representations of spin-2. Other conventions have long existed. Table \ref{tab:allcons} shows the relations of the $\rho_{M}$ parameters compared to the ad-hoc conventions known as $A_{k}$ and $\lambda_{k}$. The $\rho_{M}$ are self-explanatory because they correspond to orthonormal harmonics and transform like spin-2 representations. The arbitrary normalizations and conventions relating different basis functions have been a barrier to interpretation, needlessly complicated transformations between angular frame conventions. 

\begin{small} 

\begin{table*}[ht]
\centering

$\begin{array}{c || c | c | c | c | c}
\hline
16 \pi /3 & ( \sqrt{3}A_{0}/2 - 1/ \sqrt{3} )   & A_{1}  & A_{2}/2 & A_{5} & A_{6}\\
4\pi/(3+\lambda_{\theta})  &-\lambda_{\theta}/(3\sqrt{3}+\sqrt{3}\lambda_{\theta}) &  \lambda_{\theta\phi}/(3+\lambda_{\theta})  & \lambda_{\phi} /(3+\lambda_{\theta}) &\lambda_{\phi}^{\perp}/(3+\lambda_{\theta})  & \lambda_{\theta\phi}^{\perp} /(3+\lambda_{\theta})\\
1/c &    \rho_{0}   & \rho_{1}  & \rho_{2}  & \rho_{3}  & \rho_{4}  \\ \hline 1/(4 \pi)  &  1/\sqrt{3}-\sqrt{3} \cos^2(\theta )    & 
   \sin (2 \theta )\cos (\phi ) &  \sin ^2(\theta )\cos (2 \phi )
   & \sin^2(\theta ) \sin (2 \phi )  &  \sin (2 \theta ) \sin (\phi )
  \\   
    
   \end{array}$

  \caption{ \small Three ways of parameterizing the monopole (left of double vertical line) relative to quadrupole (right) part of the angular distribution. The bottom row (right) represents our spin-2 basis functions that are both orthogonal and uniformly normalized, compared to ad-hoc conventions of the other rows. The spin-2 coefficient combinations listed in each row are the ones that mix linearly under rotations of the frame coordinates. To find the parameterization of a given row, multiply the coefficient in each column by the function at the bottom, and add. To absolutely normalize the $A_{j}$ form, multiply the entire sum by $3/(16 \pi)$, and the normalized series will begin at $1/(4\pi)$. To absolutely normalize the $\lambda_{j}$ form multiply the entire sum by $3/(4 \pi)$. The $\rho_{M}$ form is absolutely normalized by definition. The constant $c= 3 /(8 \pi \sqrt{2})$ has been divided out to match the other conventions. The absolutely normalized form uses the sum of the $\rho_{M}$ row multiplied by $c$.}\label{tab:allcons}
\end{table*}   \end{small}

Our self-explanatory conventions for the spin-1 parameters are given in Table \ref{tab:spincon}. For example, it is quite easy to remember that $S_{x} \ra sin\theta cos \phi$, $S_{y} \ra sin\theta sin\phi$ and $S_{z}$ leads to $cos\theta$ angular dependence. 

\begin{table*}[ht]
\centering

$\begin{array}{  c | c | c  }
\hline
A_{3}& A_{7} & A_{4} \\
2A_{\phi} &2A_{\phi}^{\perp} &  2A_{\theta}  \\
S_{x} & S_{y} & S_{z} \\ \hline
 \sin (\theta )\cos(\phi)  & \sin (\theta )\sin(\phi)  & \cos(\theta)
  \\  
    
   \end{array}$

  \caption{ \small Parameterizing the spin-1 part of the angular distribution. To form the angular distribution coefficients from each row are multiplied by functions on the bottom row and added to those from Table \ref{tab:allcons}}
  \label{tab:spincon}
\end{table*}

\end{document}